\documentclass[
	aps,prl,superscriptaddress,twocolumn,
	10pt
]{revtex4-1}

\pdfoutput=1

\usepackage[final]{graphicx}
\usepackage{times,bbm,amsmath,amssymb}
\usepackage{epsfig,color}
\usepackage{xcolor}
\usepackage{hyperref}
\usepackage{float,siunitx}
\usepackage[caption = false]{subfig}
\usepackage[greek,english]{babel}
\usepackage{thumbpdf,enumerate}
\usepackage{booktabs}
\usepackage{sidecap}
\usepackage[scaled=.8]{couriers}
\usepackage{pstricks}
\usepackage{multirow}
\usepackage{placeins}
\usepackage{pst-grad,bm}
\usepackage{epigraph}
\usepackage{gensymb}
\usepackage{longtable}
\usepackage{booktabs}
\usepackage{gensymb}
\usepackage{soul}

\usepackage{acronym}
\newacro{QW}{Quantum Walk}

\usepackage{pifont}

\newcommand{\ket}[1]{\vert#1\rangle}

\newcommand{\shiftS}{\hat{\mathcal S}_{wc}}

\begin{document}
\title{Experimental engineering of arbitrary qudit states with discrete-time quantum walks} 

\author{Taira Giordani}
\affiliation{Dipartimento di Fisica, Sapienza Universit\`{a} di Roma,
Piazzale Aldo Moro 5, I-00185 Roma, Italy}

\author{Emanuele Polino}
\affiliation{Dipartimento di Fisica, Sapienza Universit\`{a} di Roma,
Piazzale Aldo Moro 5, I-00185 Roma, Italy}

\author{Sabrina Emiliani}
\affiliation{Dipartimento di Fisica, Sapienza Universit\`{a} di Roma,
Piazzale Aldo Moro 5, I-00185 Roma, Italy}

\author{Alessia Suprano}
\affiliation{Dipartimento di Fisica, Sapienza Universit\`{a} di Roma,
Piazzale Aldo Moro 5, I-00185 Roma, Italy}

\author{Luca Innocenti}
\affiliation{Centre for Theoretical Atomic, Molecular, and Optical Physics,
School of Mathematics and Physics, Queen's University Belfast, BT7 1NN Belfast, United Kingdom}

\author{Helena Majury}
\affiliation{Centre for Theoretical Atomic, Molecular, and Optical Physics,
School of Mathematics and Physics, Queen's University Belfast, BT7 1NN Belfast, United Kingdom}

\author{Lorenzo Marrucci}
\affiliation{Dipartimento di Fisica "Ettore Pancini", Universit\`{a} Federico II, Complesso Universitario di Monte Sant'Angelo, Via Cintia, 80126 Napoli, Italy}

\author{Mauro Paternostro}
\affiliation{Centre for Theoretical Atomic, Molecular, and Optical Physics,
School of Mathematics and Physics, Queen's University Belfast, BT7 1NN Belfast, United Kingdom}

\author{Alessandro Ferraro}
\affiliation{Centre for Theoretical Atomic, Molecular, and Optical Physics,
School of Mathematics and Physics, Queen's University Belfast, BT7 1NN Belfast, United Kingdom}

\author{Nicol\`o Spagnolo}
\affiliation{Dipartimento di Fisica, Sapienza Universit\`{a} di Roma,
Piazzale Aldo Moro 5, I-00185 Roma, Italy}

\author{Fabio Sciarrino}
\affiliation{Dipartimento di Fisica, Sapienza Universit\`{a} di Roma,
Piazzale Aldo Moro 5, I-00185 Roma, Italy}
\affiliation{Consiglio Nazionale delle Ricerche, Istituto dei sistemi Complessi (CNR-ISC), Via dei Taurini 19, 00185 Roma, Italy}

\begin{abstract}
The capability to generate and manipulate quantum states in high-dimensional Hilbert spaces is a crucial step for the development of quantum technologies, from quantum communication to quantum computation. One-dimensional quantum walk dynamics represents a valid tool in the task of engineering arbitrary quantum states. Here we affirm such potential in a linear-optics platform that realizes discrete-time quantum walks in the orbital angular momentum degree of freedom  of photons. Different classes of relevant qudit states in a six-dimensional space are prepared and measured, confirming the feasibility of the protocol. Our results represent a further investigation of quantum walk dynamics in photonics platforms, paving the way for the use of such a quantum state-engineering toolbox for a large range of applications.
\end{abstract}

\maketitle

\textit{Introduction ---}
The preparation of high-dimensional quantum states is of great significance in quantum information science and technology. 
Compared to qubits, \textit{qudit} states -- describing quantum systems in $d$-dimensional spaces --
enable stronger foundational tests of quantum mechanics~\cite{Vertesi2010,brunner2014bell, lapkiewicz2011experimental} and better-performing applications in secure quantum communications~\cite{bechmannpasquinucci2000quantum, fitzi2001quantum, cerf2002security, bru2002optimal, acin2003security, langford2004measuring}, quantum emulation~\cite{Buluta2009, Neeley2009}, quantum error correction~\cite{Chuang1997,Duclos-Cianci2013,Michael2016}, fault-tolerant quantum computation~\cite{bartlett2002quantum, ralph2007efficient, Lanyon2009, campbell2012magicstate, Campbell2014}, and quantum machine learning \cite{Schuld2015, Dunjko2017, Biamonte2017}. 

Protocols performed on systems living in large Hilbert spaces require great control in light of the number of parameters required to describe states and operations. Nonetheless, qudit states have been prepared successfully in various physical settings~\cite{Leibfried1996, Hofheinz2009, Neeley2009, Anderson2015, Walborn2006, Lima2011, Rossi2009, Dada2011, Anderson2015, Heeres2017, Rosenblum2018, Chu2018}. Such schemes rely on \textit{ad hoc} strategies whose dependence on the underpinning dynamics makes their translation across different physical platforms difficult. 

A promising way to achieve a higher degree of platform-universality is the use of the rich dynamics offered by \acp{QW}~\cite{aharonov1993quantum, Kempe2003, Venegas-Andraca2012}. These can be thought of as the quantum counterparts of classical random walks and comprise -- in their discrete version -- a qudit, named \emph{walker}, endowed with an internal two-dimensional degree of freedom dubbed \emph{coin}. At every time step, the walker moves coherently to neighbouring sites on a lattice, conditionally to its coin state~\cite{ambainis2001one}. \acp{QW} have been successfully implemented~\cite{Manouchehri2014} in systems as diverse as trapped atoms~\cite{cote2006quantum} and ions~\cite{Schmitz2009,Zahringer2010}, photonic circuits~\cite{Perets2008, Peruzzo2010, broome2010bdp, Schreiber2010, Rohde2011, sansoni2012quantum, boutari2016time, cardano2015quantum, cardano2016statistical, caruso2016maze}, and optical lattices~\cite{Meinert2014}. An approach for state engineering based on their dynamics offers hope of being applicable in a variety of different systems, independently of the details of the physical implementation.

While the \ac{QW} dynamics was previously shown to allow the engineering of {\it specific} walker's states~\cite{chandrashekar2008optimizing,majury2016robust}, in Ref.~\cite{Innocenti2017} a scheme was proposed to use discrete-time \acp{QW} on a line to prepare {\it arbitrary} qudit states with high probability. This is achieved by enhancing the degree of control over the walk's dynamics through the arrangement of suitable step-dependent \emph{coin} operations, which affect the coin-walker quantum correlations by \emph{de facto} steering the state of the walker towards the desired final state, and finally projecting in the coin space. This removes the correlations between walker and coin, thus producing a pure walker state with the desired features. In light of the large parameter space that characterizes the problem at hand, a systematic approach to the identification of the right set of coin operations and final projection is necessary.

\begin{figure*}[t!]
\includegraphics[width=\textwidth]{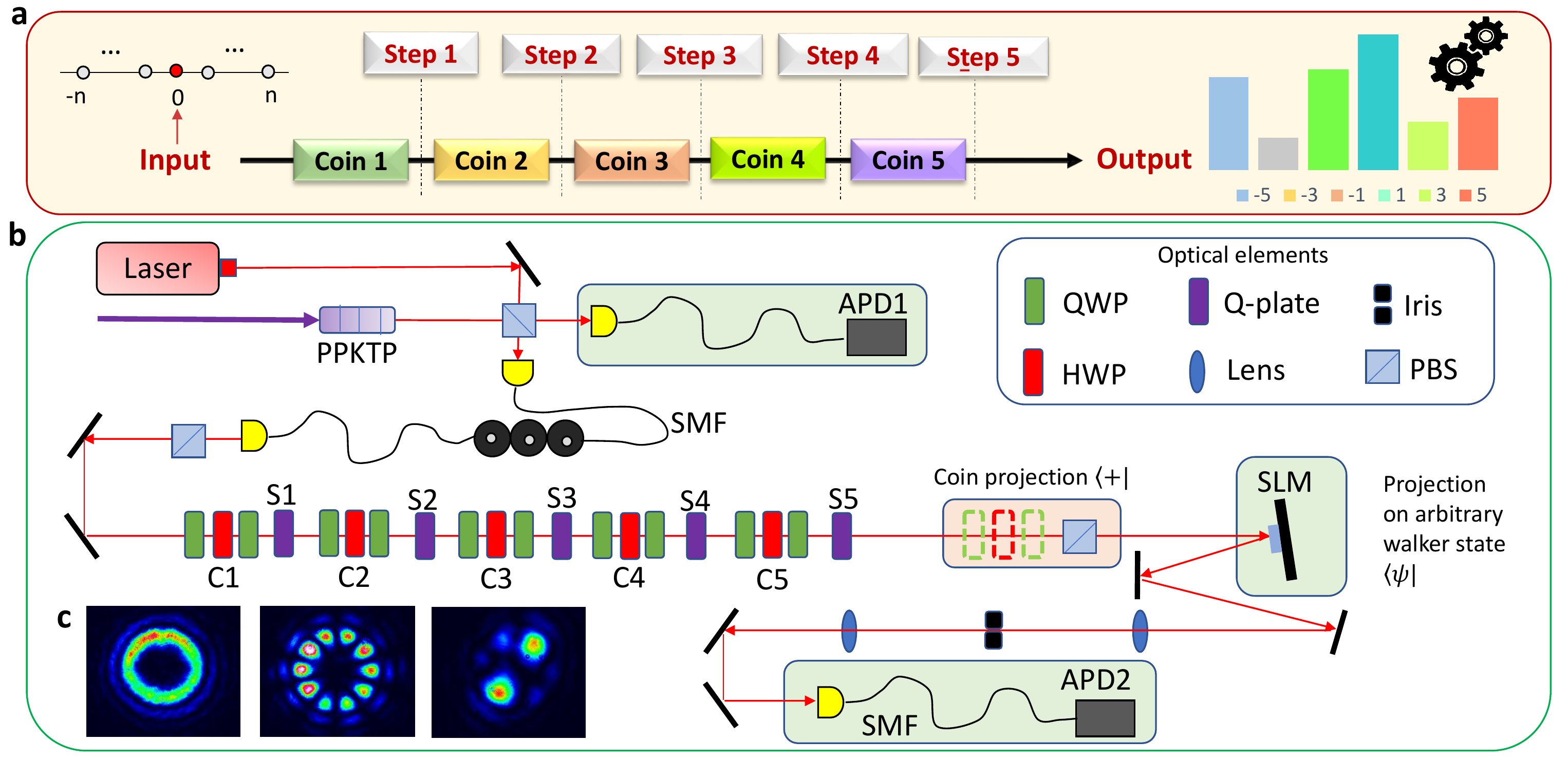}
\caption{Set up for the quantum state engineering toolbox. {\bf a)} Conceptual scheme of the protocol. At each step of \ac{QW} the coin operator is changed to obtain a target state in the output. {\bf b)} A single-photon source, composed of a periodically-poled potassium titanyl phosphate (PPKTP), generates pairs of photons that are coupled in a single-mode fibre (SMF). One photon acts as trigger while the other is prepared in $\ket{\psi_0}=\ket{+}\otimes 
\ket{0}$ through polarization controllers and a polarizing beam splitter (PBS). Five sets of quarter (QWP) and half (HWP) waveplates implement the operators $\{ C_i\}$ for each step. Five Q-plates (QP) implement the shift operator of the \ac{QW}
$\{S_i\}$. The detection stage consists of a PBS followed by a spatial light modulator (SLM), 
a SMF and an avalanche photodiode detector (APD), for the projection onto $\ket{+} \otimes \ket{\psi}$. {\bf c)} Pictures of OAM modes of the output states after PBS, obtained with coherent light. From right: OAM eigenstate corresponding to $m=5$; balanced superposition of $m=\pm 5$; balanced superposition of all OAM components covered by 5-step \ac{QW} $m=\{\pm 5, \pm 3, \pm 1\}$. }
\label{app}
\end{figure*}

In this paper, we use of the scheme of Ref.~\cite{Innocenti2017} to demonstrate a state-engineering protocol based on the controlled dynamics generated by \acp{QW}. We use the orbital angular momentum (OAM) degree of freedom of single-photon states as a convenient embodiment of the walker~\cite{zhang-oam-qw-2010,goyal2013implementing,cardano2015quantum}. OAM-based experiments offer the possibility to cover Hilbert spaces of large dimensions in light of the favourable (linear) scaling of the number of optical elements with the size of the walk. Moreover, the scheme allows for the full control of the coin operation that is key to the implementation of the walk. In order to demonstrate the versatility of our scheme, we focus on the interesting classes of cat-like states and spin-coherent states~\cite{brune_Cat_1992,monroe_Cat_1996}. Furthermore, we show experimentally the capability of engineering arbitrary states.
The quality of the generated states and the feasibility of the experimental protocol that we have put in place, demonstrate the effectiveness of a hybrid platform for quantum state engineering. Such platform holds together a programmable quantum system, the photonic \ac{QW} in the angular momentum, and classical optimization algorithms to effectively reach a given target. 

\textit{Engineering quantum walks.--}
We consider a discrete-time \ac{QW} with a two-dimensional coin with logical states labelled as $\{\ket{{\downarrow}}_c, \ket{{\uparrow}}_c\}$. The dynamics are made up of consecutive unitary steps. At step $t$, a \emph{coin operator} $\hat{\mathcal{C}_{t}}$ changes the coin state and is then followed by a \emph{shift operator} $\shiftS$, which moves the walker conditionally to the coin state. Such transformations are described by the operators
\begin{equation}
\hat{\mathcal{C}_t}=
\left(
\begin{array}{ll}
e^{i \xi_t} \cos{\theta_t} &  e^{i \zeta_t} \sin{\theta_t} \\
-e^{-i \zeta_t} \sin{\theta_t} & e^{-i \xi_t} \cos{\theta_t}
\end{array}
\right),
\label{coinExpr}
\end{equation}
which accounts for the coin tossing, and
$\shiftS=\sum_k |k-1\rangle \langle k|_w\otimes |{\downarrow}\rangle \langle {\downarrow}|_c+ |k+1\rangle \langle k|_w\otimes |{\uparrow}\rangle \langle{\uparrow}|_c$,
which realizes the conditional motion of the walker. Here $k$ is the lattice-site occupied by the walker and $\{\theta_t,\xi_t,\zeta_t \}$ are parameters identifying a unitary transformation in two dimensions. The evolution through $n$ steps of the {QW} is given by $\hat{U}=\prod_{t=1}^n \shiftS\hat{\mathcal{C}_t}$.

\begin{figure*}[t!]
\includegraphics[width=\textwidth]{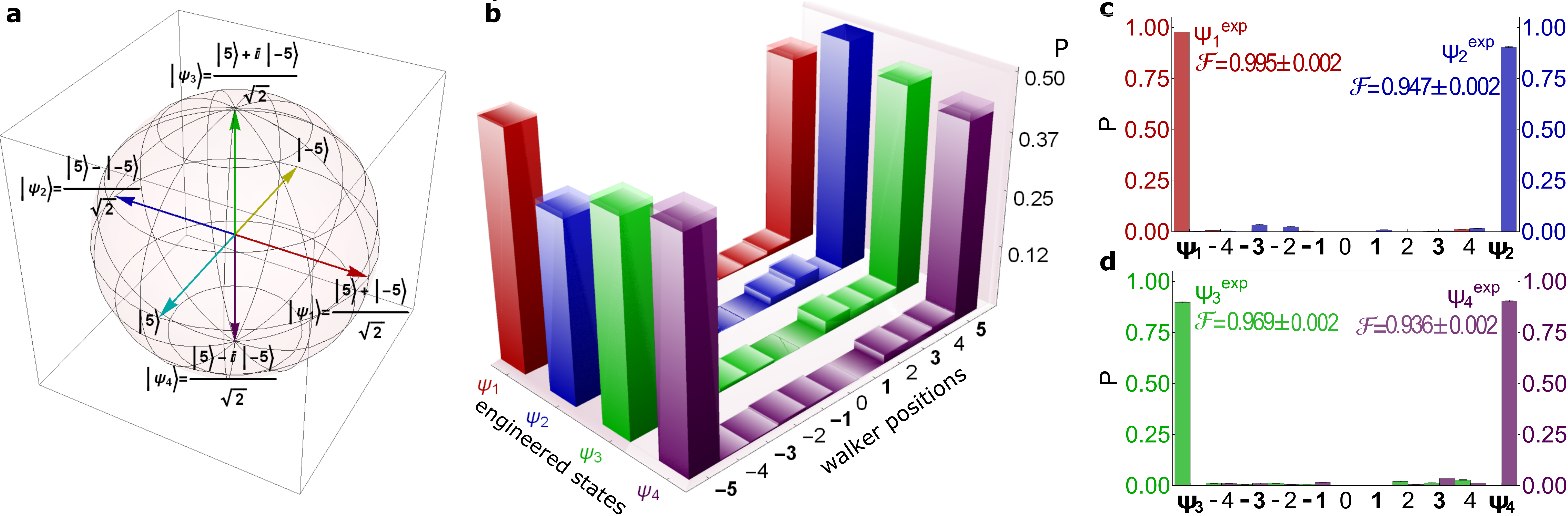}
\caption{
	Experimental results for the engineering of angular momentum cat states. {\bf a)} Representation on a Bloch-like ball of the four target states corresponding to the superposition of $\ket{\pm 5}$, which correspond to OAM states with maximum and minimum projection of the angular momentum along the quantization axis. {\bf b)} Population of the OAM components after 5-step {QW} for the states $\ket{\psi_i}~(i=1,2,3,4)$ in panel {\bf a)}. Odd-$m$ position states (bold numbers on $x$-axis) should be the only ones involved in the state engineering. However, we report also the populations of even-$m$ position states (light-black numbers on $x$-axis) to illustrate possible imperfections at the generation and detection stages. The error bars associated with the experimental populations are shown by the transparent areas on top of each histogram. {\bf c)}-{\bf d)} Distributions of the probabilities $P_i=\langle B^{(j)}_i\vert\rho_\text{exp}\vert B^{(j)}_i\rangle~(j=1,2)$ that the experimental walker state $\rho_\text{exp}$ is found to be one of the elements of the bases $B^{(j)}=\{\ket{\psi_p},\ket{\psi_{p+1}},\ket{\pm4},\ket{\pm3},\ket{\pm2},\ket{\pm1},\ket{0}\}$ with $p=1$ for $j=1$ and $p=3$ for $j=2$. All the error bars are due to Poissonian uncertainties, propagated through Monte Carlo methods. The state fidelities ${\cal F}$ are calculated as described in the main text.
}
\label{fig55}
\end{figure*}

In Ref.~\cite{Innocenti2017} it was shown that it is always possible to find a set of coin operators $\{\hat{\mathcal{C}_t}\}_{t=1}^{n}$ that produce an arbitrary target state in the full coin-walker space. In addition, via suitable projection in the coin space, arbitrary walker states can also be obtained. The identification of the correct set of coin operators is enabled by a classical algorithm to maximize the fidelity between the final state of the walker, after projection of the coin, and the target $(n+1)$-dimensional state.

To demonstrate the effectiveness of this approach for the state engineering of high-dimensional spaces, we here focus on classes of physically relevant states. First, we consider the synthesis of angular-momentum Schr\"odinger cat states~\cite{AMcat}, achieved by engineering coherent superpositions of extremal walker positions. The correspondence between the position space of the walker and an angular momentum of quantum number $n/2$, which will be illustrated and clarified later in this paper, makes the \ac{QW} perfectly suited to synthesize this class of states. Schr\"odinger cat states play a crucial role in the investigations on foundations of quantum mechanics~\cite{schrodingerCAT} and their generation is at the core of various quantum engineering protocols~\cite{brune_Cat_1992, monroe_Cat_1996,agarwalCAT1997, zhang2016creating}. The second class of states that we consider is \emph{spin-coherent states}~\cite{ulyanov_spin1999}, which are the spin-like counterpart of \emph{coherent states} of a quantum harmonic oscillator. Finally, in order to validate the flexibility of our approach, we demonstrate high-quality engineering of both balanced, and randomly sampled states.

\textit{Experimental apparatus ---}
We have implemented a discrete-time \ac{QW} with $n=5$ steps, using the angular momentum states of light $\{|m\rangle_w\}$ ($m=\pm 5, \pm 3, \pm 1$) as the physical embodiment of the walker, while the logical states of the coin are encoded in circular-polarization states $\{|R\rangle, |L\rangle\}$. We dub such degree of freedom as {\it spin angular momentum} (SAM) to mark the difference with OAM. Our experimental setup, which is shown schematically in Fig.~\ref{app} and follows Refs.~\cite{cardano2015quantum,cardano2016statistical}, allows for the full coin-walk evolution to take place in a single light beam, thus avoiding an exponential growth of optical paths as in previous interferometric implementations~\cite{zhang-oam-qw-2010,goyal2013implementing,cardano2015quantum}. Arbitrary coin operators are achieved through a sequence of suitably arranged and oriented quarter- and half-waveplates~\cite{simonMukunda}. The shift operator $\shiftS$ is instead implemented using a \emph{Q-plate} (QP)~\citep{marrucci-2006spin-to-orbital}, an active device that uses an inhomogeneous birefringent medium to convert SAM into OAM  and that can conditionally change the values of the OAM by a quantity $2q$ (here $q$ is the topological charge of the device) according to transformations 
\begin{equation}
\begin{aligned}
|L,m\rangle&\stackrel{\text{QP}}{\longrightarrow} \cos{ \frac{\delta}{2}}|L,m\rangle + i e^{2 i \alpha_0}\sin{\frac{\delta}{2}}|R,m+2q\rangle,\\
|R,m\rangle&\stackrel{\text{QP}}{\longrightarrow} \cos{ \frac{\delta}{2}}|R,m\rangle + i e^{-2 i \alpha_0}\sin{\frac{\delta}{2}}|L,m-2q\rangle.
\end{aligned}
\end{equation}
The additional phase $\alpha_0$ between the two polarizations is compensated by changing the orientations of the waveplates which implement the coin operator of the subsequent step.

\begin{figure*}[t]
\includegraphics[width=\textwidth]{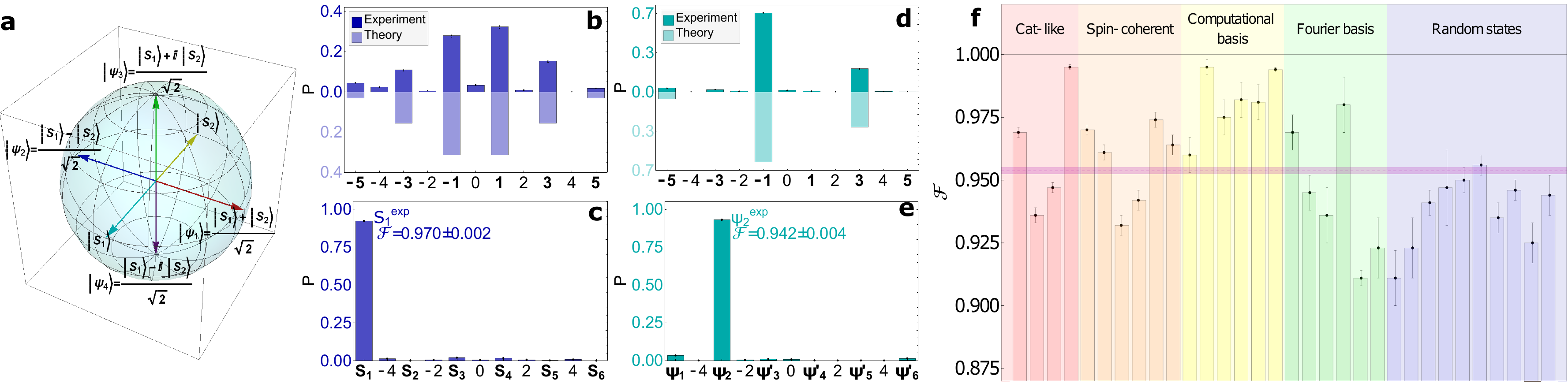}
\caption{Experimental results for the engineering of SCSs and their coherent superposition: {\bf a)} Bloch-sphere representation for the mutually orthogonal SCSs $\ket{S_1}$ and $\ket{S_2}$.
{\bf b)} Probability distributions associated to the projection of $\ket{S_1}$ onto the computational basis. As previously explained, we also consider the contribution of even OAM components.
{\bf c)} Probability distribution corresponding to the basis that contains the target state itself $\ket{S_1}$, generated with the fidelity reported in the panel. Such orthonormal basis correspond to eigenstates of $S_x$ for a particle with $s=5/2$.
{\bf d)} Experimental probability distribution on computational basis for $\ket{\psi_2}=\frac{1}{\sqrt{2}}\left (\ket{S_1}- \ket{S_2} \right)$. Only components $\{-5, -1, 3\}$, corresponding to the logical states $\{1,3,5\}$, have non-zero probabilities.
{\bf e)} Quantum state fidelity evaluated measuring the state $\ket{\psi_2}$ on the orthonormal basis that contains the state $\ket{\psi_1}$, as described in the main text. 
{\bf f)} Summary of quantum state fidelities for the $32$ states generated in the experiment. The average fidelity, $\bar{\mathcal{F}}=0.954\pm 0.001$, is reported by the magenta area. 
}
\label{figSpin}
\end{figure*}

Single-photon states are generated via a type-II, collinear spontaneous-parametric-down-conversion source [cf. Fig.~\ref{app}]. 
The photons emitted by the source are separated with a polarizing beam splitter (PBS) and coupled to two single-mode fibers (SMF). One photon acts as the trigger signal, while the other one undergoes the \ac{QW} evolution. After the propagation in the SMF and the first PBS, the initial state of the walker and coin is prepared in $\ket{\psi_0}_{wc}=\ket{0}_w \otimes \ket{+}_c$ with $\ket{+}_c=(\ket{{\uparrow}}_c+\ket{{\downarrow}}_c)/\sqrt2$. At the end of an $n$-step \ac{QW}, the protocol involves a projection of the coin state onto $|+\rangle_c$. This is experimentally implemented by a final PBS. The OAM analysis is performed through a \emph{spatial-light modulator} (SLM) followed by coupling into a single-mode fiber, which allows for the measurement of arbitrary superposition of OAM components with high accuracy~\cite{bolduc2013holo,dambrosio_mubs2013}. The quantum state fidelity between the actual state of the walker and the target $(n+1)$-dimensional state is estimated by projecting the OAM state onto a basis that contains the given target state [cf. Fig.\ref{app}]. 

\textit{Engineering cat-like states in high dimensions.--} Our investigation on the engineering of quantum states living in Hilbert spaces of large dimensions starts from coherent superpositions of two extremal lattice sites of the walker. The isomorphism of the OAM with an angular momentum of quantum number $n/2$ allows us to put in correspondence the position states of the walker on the lattice $\ket{\pm 5}$ with angular momentum states with minimum and maximum projections onto the quantization axis $\ket{\pm 5/2}$ (for simplicity of notation, we will use position states only). Such isomorphism makes a coherent superposition state such as $(\ket{5}+e^{i\varphi}\ket{-5})/\sqrt{2}$ (with $\varphi$ a suitable phase) a faithful angular momentum Schr\"odinger cat state~\cite{AMcat}, thus benchmarking the performance of our experiment with a relevant class of states~\cite{chandrashekar2008optimizing,zhang2016creating,majury2016robust} that is also used in quantum sensing~\cite{zeilengerOAM,dambrosio_gear2013}.

In Fig.~\ref{fig55} we report the experimental results for the generation of four of such states, which are conveniently pictured as the states pointing towards the poles of a Bloch-like ball. Quantum coherence between the components of such states has been tested by changing their relative phase. The values of the state fidelity between the experimentally synthesized states and their respective target ones are reported in Fig.~\ref{fig55}. Hereafter we compute fidelities by projecting the state on the orthonormal basis which includes the target qudit in the 6-dimensional subspace associated to our 5-step \ac{QW}, generated by the OAM eigenstate  $\{|m\rangle_w\}$ ($m=\pm 5, \pm 3, \pm 1$).

The second class of relevant states that we addressed are \emph{spin-coherent states} (SCSs)~\cite{agarwalCAT1997}. These are the counterparts of coherent states of the harmonic oscillator for a particle with spin $s$~\cite{radcliffeSpin,arecchiSpinCoherent,agarwalCAT1997,vedralSpin}. 
SCSs are eigenstates -- with eigenvalue $s$ -- of the component of the total spin-momentum operator $\hat{S}$ pointing along the direction identified by the polar spherical angles $\{ \theta, \phi \}$~\cite{arecchiSpinCoherent,agarwalCAT1997,ulyanov_spin1999,yenLee_spin2015} 
A decomposition of such states over the $\{\ket{s_z}\}$ basis of the projected spin along z-direction ($\hat{S}_z$) reads
\begin{equation}
\begin{aligned}
 \ket{s,\theta,\phi} &= \sum_{s_z=-s}^{s}\sqrt{\frac{(2s)!}{(s+s_z)!(s-s_z)!}} e^{-i\phi s_z} C_\theta^{s+s_z}
S_\theta^{s-s_z}\ket{s_z}
 \label{spin_coherent1}
\end{aligned}
   \end{equation}
with $C_\theta=\sqrt{1-S^2_\theta}=\cos(\theta/2)$. SCSs have various applications in condensed matter physics, in particular in quasi-exactly solvable models, for Wigner-Kirkwood expansion and in quantum correction to energy quantization rules~\cite{ulyanov_spin1999}. At the foundational level, they can be used very fruitfully to generate Schr{\"o}dinger cat states~~\cite{agarwalCAT1997}. 

Although SCSs are in general not orthogonal, they form a convenient basis. Moreover, as two SCSs pointing in opposite azimuthal directions are orthogonal for $\theta\sim\pi/2$, by restricting the attention to $\{\ket{s,\pi/2,\phi},\ket{s,-\pi/2,\phi}\}$ we would be dealing with an orthonormal basis, which we can use to construct the analogous of a Bloch ball for a two-level system (cf. Fig.~\ref{figSpin}a). We have thus engineered $\ket{S_1}\equiv\ket{5/2,\pi/2,0}$ and $\ket{S_2}\equiv\ket{5/2,-\pi/2,0}$, and considered the experimental synthesis of balanced coherent superpositions of such states. Such superpositions are akin to the Schr\"odinger cat states built on coherent states of a harmonic oscillator, as they exhibit signatures of non-classical interference~\cite{agarwalCAT1997,SI}. For instance, only even (odd) components of the logical basis enter the superposition $\ket{S_1}+\ket{S_2}$ ($\ket{S_1}-\ket{S_2}$), a parity rule that is fully analogous to the one characterizing even (odd) bosonic cat states. Thanks to the isomorphism between the spaces of OAM and of arbitrary angular momentum equal to $n/2$, we can generate SCS mapping the basis $\{ \ket{s_z}\}$ in (\ref{spin_coherent1}) into the basis of the \ac{QW} $\{|m\rangle_w\}$. The results are illustrated in Fig.~\ref{figSpin}a-e, where we show the high quality of both the SCSs and SCS-based cat states that we have generated. 

\emph{Engineering arbitrary qudits.--} In order to demonstrate the flexibility of our scheme, we have addressed the generation of states of arbitrary complexity, starting from balanced states and then moving towards randomly chosen states. Balanced states are challenging as one needs to ensure equal population of all their components, a condition that is very prone to experimental imperfections. Assessing the quality of generation of such states thus provides a significant benchmark to the effectiveness of the employed procedure. We have then engineered the element of a Fourier basis associated to the Hilbert space of the walker. This choice is motivated by the importance of quantum Fourier transform in quantum algorithms \cite{chuang2010}, as well as its role in the identification of mutually unbiased bases for quantum cryptography and communication in high-dimensions~\cite{durtMUB,VatanMUBS,brierley_mubs2013,dambrosio_mubs2013}.

Final measurements concern the generation of randomly-chosen qudits. We have engineered up to 5 states with real-valued amplitudes and 5 with complex-valued ones, where the state components are sampled from a uniform distribution (cf. Ref.~\cite{SI}).
In Fig.~\ref{figSpin}f quantum state fidelities are reported for all the experimental engineered states, including the Fourier Basis and random sampled qudits, where the red area shows the average fidelity and its uncertainty ($\mathcal{F}{=}0.954 \pm 0.001$)~\cite{SI}. Such test provides a further proof of the effectiveness of the strategy demonstrated in our experiment.

\textit{Discussion.--} We have successfully tested a QW-based quantum state engineering strategy assisted by numerical optimization~\cite{Innocenti2017}. Our tests have been run in a photonic platform using OAM as the embodiment of a quantum walker. This allowed us to implement a five-step QW, without exponential overhead in the number of required optical paths and with full control on the preparation, coin-operation, and detection stages. We focused on significant instances of high-dimensional states to benchmark the effectiveness of the protocol, demonstrating its ability to synthesize high-quality cat-like states. Our results show the viability of \ac{QW}-based approaches to state engineering, reinforcing the idea that numerical optimization complementing quantum dynamics of a sufficient degree of complexity is effective for high-dimensional state engineering. Further improvements of our approach can be envisaged by identifying appropriate routines to optimize the state engineering process in the presence of actual experimental imperfections. To this end, machine learning algorithms can be a promising add-on to our numerical optimization approach to adapt the coin operators to a given experimental implementation.

\begin{acknowledgments}
\textit{Acknowledgements.--}
We acknowledge support from the ERC-Advanced grant PHOSPhOR (Photonics of Spin-Orbit Optical Phenomena; Grant Agreement No. 694683), the Northern Ireland Department for Economy (NI DfE), the DfE-SFI Investigator Programme (grant 15/IA/2864), and the H2020 Collaborative Project TEQ (grant 766900). 
\end{acknowledgments}

\end{document}


\title{Supplementary material: Experimental engineering of arbitrary qudit states with discrete-time quantum walks} 

\author{Taira Giordani}
\affiliation{Dipartimento di Fisica, Sapienza Universit\`{a} di Roma,
Piazzale Aldo Moro 5, I-00185 Roma, Italy}

\author{Emanuele Polino}
\affiliation{Dipartimento di Fisica, Sapienza Universit\`{a} di Roma,
Piazzale Aldo Moro 5, I-00185 Roma, Italy}

\author{Sabrina Emiliani}
\affiliation{Dipartimento di Fisica, Sapienza Universit\`{a} di Roma,
Piazzale Aldo Moro 5, I-00185 Roma, Italy}

\author{Alessia Suprano}
\affiliation{Dipartimento di Fisica, Sapienza Universit\`{a} di Roma,
Piazzale Aldo Moro 5, I-00185 Roma, Italy}

\author{Lorenzo Marrucci}
\affiliation{Dipartimento di Fisica "Ettore Pancini", Universit\'a Federico II, Complesso Universitario di Monte Sant’Angelo, Via Cintia, 80126 Napoli, Italy}

\author{Luca Innocenti}
\affiliation{Centre for Theoretical Atomic, Molecular, and Optical Physics,
School of Mathematics and Physics, Queen’s University Belfast, BT7 1NN Belfast, United Kingdom}

\author{Helena Majury}
\affiliation{Centre for Theoretical Atomic, Molecular, and Optical Physics,
School of Mathematics and Physics, Queen's University Belfast, BT7 1NN Belfast, United Kingdom}

\author{Mauro Paternostro}
\affiliation{Centre for Theoretical Atomic, Molecular, and Optical Physics,
School of Mathematics and Physics, Queen's University Belfast, BT7 1NN Belfast, United Kingdom}
\affiliation{Laboratoire Kastler Brossel, ENS-PSL Research University, 24 rue Lhomond, F-75005 Paris, France}

\author{Alessandro Ferraro}
\affiliation{Centre for Theoretical Atomic, Molecular, and Optical Physics,
School of Mathematics and Physics, Queen's University Belfast, BT7 1NN Belfast, United Kingdom}

\author{Nicol\'o Spagnolo}
\affiliation{Dipartimento di Fisica, Sapienza Universit\`{a} di Roma,
Piazzale Aldo Moro 5, I-00185 Roma, Italy}

\author{Fabio Sciarrino}
\affiliation{Dipartimento di Fisica, Sapienza Universit\`{a} di Roma,
Piazzale Aldo Moro 5, I-00185 Roma, Italy}
\affiliation{Consiglio Nazionale delle Ricerche, Istituto dei sistemi Complessi (CNR-ISC), Via dei Taurini 19, 00185 Roma, Italy}
\maketitle

\section{Summary of the states engineered}
In the following table we report the summary of target states engineered during the experiment, with relative quantum state fidelities and generation probabilities. The latter is provided by the algorithm developed in Ref.\cite{Innocenti2017} together by the coin operators needed in the engineering process. The expected fidelities for all states is 1. The protocol and the experimental platforms are tested firstly with trivial states, as the element of computational basis corresponding to the eigenstate of OAM operator $\{\ket{m} \}=\{\ket{\pm5}, \ket{\pm3},\ket{\pm1} \}$. Then, superposition of two OAM components up to more complex states with arbitrary no-zero amplitudes on OAM basis, such as spin-coherent states, the Fourier basis and random extracted states. Quantum state fidelities are calculated measuring target state on orthonormal basis which contains the state itself. They are constructed according to Gram-Schmidt orthogonalization, starting from an ensemble of linearly independent states composed by elements of the computational basis and the target state. 

Let us clarify the notation used in Table\ref{table1}. For the Fourier basis the convention employed is the following: $\ket{QFT_k}=\frac{1}{\sqrt{6}}\sum_{j=1}^{6}e^{\frac{ i \pi jk }{3}}\ket{j}$, where $\{\ket{j}\}$ stands for the logical basis that in our case corresponds to the OAM eigenstates $\{\ket{m} \}$. The notation $\ket{r_k}$ and $\ket{c_k}$ refers to real and complex random states respectively. Amplitudes of real states have been sampled uniformly in the range $\left[0,1\right]$ and then normalized. In the case of complex states we have sampled the real and imaginary part separately in the range $\left[-0.5,0.5\right]$. In Table\ref{table2} we report the resulting amplitudes.
\begin{table*}[h!]
\centering
\begin{tabular}{lcc|lcc}
\toprule
Target State & $Probability$  & $\mathcal{F}_{exp}$  & Target State & $Probability$  & $\mathcal{F}_{exp}$ \\
\midrule
 $\quad\ket{-5}$ & $0.5$  & $0.981 \pm 0.007\quad$ &$\quad\ket{QFT_1}\qquad $ & $0.14$ & $0.969\pm 0.007$\\
 $\quad\ket{-3}$ & $0.5$ & $0.982 \pm 0.007\quad$ &$\quad\ket{QFT_2}$ & $0.17$ & $0.923\pm 0.022$ \\
 $\quad\ket{-1}$ & $0.5$ & $0.960 \pm 0.007\quad$ &$\quad\ket{QFT_3}$ & $0.17$ & $0.911\pm 0.011$\\ 
 $\quad\ket{1}$ & $0.5$ & $0.995 \pm 0.007\quad$ & $\quad\ket{QFT_4}$&$0.17$ &  $0.980\pm 0.011$ \\
 $\quad\ket{3}$ & $0.5$ & $0.975 \pm 0.007\quad$ & $\quad\ket{QFT_5 }$& $0.17$& $0.936\pm 0.011$ \\
 $\quad\ket{5}$ & $0.5$ & $0.994 \pm 0.001\quad$ & $\quad\ket{QFT_6} $& $0.17$ & $0.945\pm 0.007$ \\
 $\quad\frac{1}{\sqrt{2}}\left(\ket{-5}+\ket{5} \right)$ & $0.5$  & $0.995 \pm 0.001\quad$ &$\quad\ket{r_1}$ & 0.22& $0.911\pm0.011$\\
 $\quad\frac{1}{\sqrt{2}}\left(\ket{-5}-\ket{5}\right)$ & $0.5$ & $0.947 \pm 0.002\quad$ &$\quad\ket{r_2}$  & 0.16 & $0.923 \pm 0.012$\\
 $\quad\frac{1}{\sqrt{2}}\left(\ket{-5}+i\ket{5}\right)$ & $0.5$ & $0.969 \pm 0.002\quad$ &$\quad\ket{r_3}$  & 0.17 & $0.941 \pm 0.004$\\ 
  $\quad\frac{1}{\sqrt{2}}\left(\ket{-5}-i\ket{5}\right)$ & $0.5$ & $0.936 \pm 0.003\quad$&$\quad\ket{r_4} $& 0.14 &$0.947 \pm 0.015$  \\
 $\quad\ket{S_1}=\ket{5/2,\pi /2,0}$   & $0.15$ & $0.970 \pm 0.002\quad$ &$\quad\ket{r_5}$ & 0.19 & $0.950\pm0.005$  \\
 $\quad\ket{S_2}=\ket{5/2,-\pi /2,0}$ & $0.15$ & $0.961 \pm 0.003\quad$ &$\quad\ket{c_1}$ &0.16 & $0.956\pm0.004$ \\
 $\quad\frac{1}{\sqrt{2}}\left(\ket{S_1}+\ket{S_2}\right)$ & $0.15$ & $0.932 \pm 0.004\quad$&$\quad\ket{c_2}$ & 0.29 &$0.935 \pm 0.006$  \\
 $\quad\frac{1}{\sqrt{2}}\left(\ket{S_1}-\ket{S_2}\right)$ & $0.15$ & $0.942 \pm 0.004\quad$& $\quad\ket{c_3}$& 0.17 & $0.925\pm 0.008$ \\
$\quad\frac{1}{\sqrt{2}}\left(\ket{S_1}-i\ket{S_2}\right)$ & $0.23$ & $0.974 \pm 0.003\quad$&$\quad\ket{c_4}$ & 0.16 & $0.944\pm0.008$\\
$\quad\frac{1}{\sqrt{2}}\left(\ket{S_1}+i\ket{S_2}\right)$ & $0.23$ & $0.964 \pm 0.004\quad$& $\quad\ket{c_5}$ & 0.28 & $0.946\pm0.004$\\
\bottomrule
\end{tabular}
\caption{Summary of the measured states with relative generation probabilities and experimental quantum state fidelities. }
\label{table1}
\end{table*}

\begin{table*}[bh!]
\centering
\begin{tabular}{cc}
\toprule
State & Amplitudes \\
\midrule
$\qquad\ket{r_1}\qquad$& $\left( 0.51, 0.27, 0.13, 0.10, 0.29, 0.75\right)$\\
$\qquad\ket{r_2}\qquad$& $\left( 0.19, 0.40, 0.04, 0.53, 0.37, 0.62\right)$\\
$\qquad\ket{r_3}\qquad$&$\left( 0.50, 0.74, 0.40, 0.16, 0.10, 0.006\right)$ \\
$\qquad\ket{r_4}\qquad$& $\left( 0.50, 0.47, 0.55, 0.31, 0.36, 0.04\right)$ \\
$\qquad\ket{r_5}\qquad$& $\left( 0.24, 0.12, 0.72, 0.16, 0.54, 0.30\right)$ \\
$\qquad\ket{c_1}\qquad$& $\left( 0.04+0.35i, 0.34+0.41i, 0.10+0.42i, 0.18-0.26i, 0.11-0.11i, -0.47+0.22i\right)$  \\
$\qquad\ket{c_2}\qquad$& $\left( 0.19-0.33i, -0.43+0.30i, -0.18-0.02i, -0.37+0.42i, -0.12-0.10i, 0.23+0.38i\right)$\\
$\qquad\ket{c_3}\qquad$& $\left( -0.19-0.30i, -0.02+0.39i, 0.30-0.15i, 0.25-0.22i, -0.13+0.42i, 0.24+0.48i\right)$\\
$\qquad\ket{c_4}\qquad$& $\left( 0.06+0.07i, 0.30-0.37i, -0.23+0.08i, 0.11-0.13i, -0.22+0.57i, 0.07-0.54i\right)$\\
$\qquad\ket{c_5}\qquad$& $\left( 0.07+0.14i, 0.48-0.34i, -0.41-0.18i, -0.41-0.09i, -0.10+0.32i, 0.32+0.18i\right)$\\
\bottomrule
\end{tabular}
\caption{Amplitudes of random states.}
\label{table2}
\end{table*}

\section{Cat states based on spin coherent states: phase-space picture}

In the main manuscript we have introduced the decomposition of a spin coherent state (SCS) $\ket{s,\theta,\phi}$ over the basis of eigenstates of angular momentum $\{\ket{s_z}\}$. This reads
\begin{equation}
\ket{s,\theta,\phi}=\sum^s_{s_z=-s}\sqrt{\frac{(2s)!}{(s+s_z)!(s-s_z)!}}e^{-i\phi s_z}C^{s+s_z}_\theta S^{s-s_z}_\theta\ket{s_z},
\end{equation}
where the functions $C_{\theta}$ and $S_\theta$ have been defined in the main manuscript. We have also introduced the SCS-based Schr{\"o}dinger cat states built as the following superpositions of orthogonal states $\ket{S_1}:=\ket{5/2,\pi/2,0}$ and $\ket{S_2}:=\ket{5/2,-\pi/2,0}$:
\begin{equation}
\ket{\psi_{1}}=\frac{1}{\sqrt 2}(\ket{S_1}+\ket{S_2}),\qquad\ket{\psi_{2}}=\frac{1}{\sqrt 2}(\ket{S_1}-\ket{S_2}).
\end{equation}
In this Section, we aim at providing a brief analysis of the features of such states, which are best analyzed in a suitably defined phase space~\cite{agarwalCAT1997}. In particular, we shall be considering the analogous of the Husimi $Q$ function~\cite{WM} defined as
\begin{equation}
\label{deco}
Q_j(\alpha,\beta)=\vert\langle{5/2,\alpha,\beta}\vert\psi_j\rangle\vert^2\qquad(j=1,2)
\end{equation}
in the spherical polar space where the Cartesian coordinates $(x,y,z)$ are mapped into $x\to Q_j(\alpha,\beta)\sin\alpha\cos\beta$, $y\to Q_j(\alpha,\beta)\sin\alpha\sin\beta$ and $z\to Q_j(\alpha,\beta)\cos\alpha$. Despite the simplicity of its definition, $Q_j(\alpha,\beta)$ captures important information about the quantum interference between the orthogonal components of $\ket{\psi_{j}}$, which differentiate such states from the incoherent mixture of SCSs $(\ket{S_1}\bra{S_1}\pm\ket{S_2}\bra{S_2})/2$. 

The orthogonality of $\ket{S_1}$ and $\ket{S_2}$ allows one to cast $Q_j(\alpha,\beta)$ as 
\begin{equation}
Q_j(\alpha,\beta)=\frac12\left(|q_+(5/2,\alpha,\beta)|^2+|q_-(5/2,\alpha,\beta)|^2+\text{sign}_j2\text{Re}[q_+(5/2,\alpha,\beta)q^*_-(5/2,\alpha,\beta)]\right)
\end{equation}
where $q_\pm(s,\alpha,\beta)=\bra{s,\alpha,\beta}{s,\pm\theta,0}\rangle$ and $\text{sign}_1=-\text{sign}_2=+1$. Such scalar products can be evaluated explicitly for any value of $s$ by using the decomposition in Eq.~\eqref{deco} to get 
\begin{equation}
\begin{aligned}
q_\pm(\alpha,\beta)&=(\pm 1)^s\frac{\Gamma (2 s+1)}{\Gamma (s+1)^2} S^s\left({\alpha }\right) C^s\left(\alpha\right) S^s\left(\theta\right) C^s\left({\theta
   }\right) \left[\, _2F_1\left(1,-s;s+1;\mp e^{-i \beta  } T\left({\alpha}\right) T\left({\theta }\right)\right)\right.\\
  &+\left.{}_2F_1\left(1,-s;s+1;\mp e^{i \beta} T^{-1} \left({\alpha}\right) T^{-1}\left({\theta }\right)\right)-1\right],
  \end{aligned}
\end{equation}
where $T(\alpha)=S(\alpha)/C(\alpha)=\tan(\alpha/2)$, $_2F_1(a,b,c;d)$ is the ordinary Hypergeometric function, and $\Gamma(d)$ is the Gamma function with argument $d$. 
\begin{figure}[!t]
\centering
{\bf (a)}\hskip8cm{\bf (b)}\\
\includegraphics[width=\columnwidth]{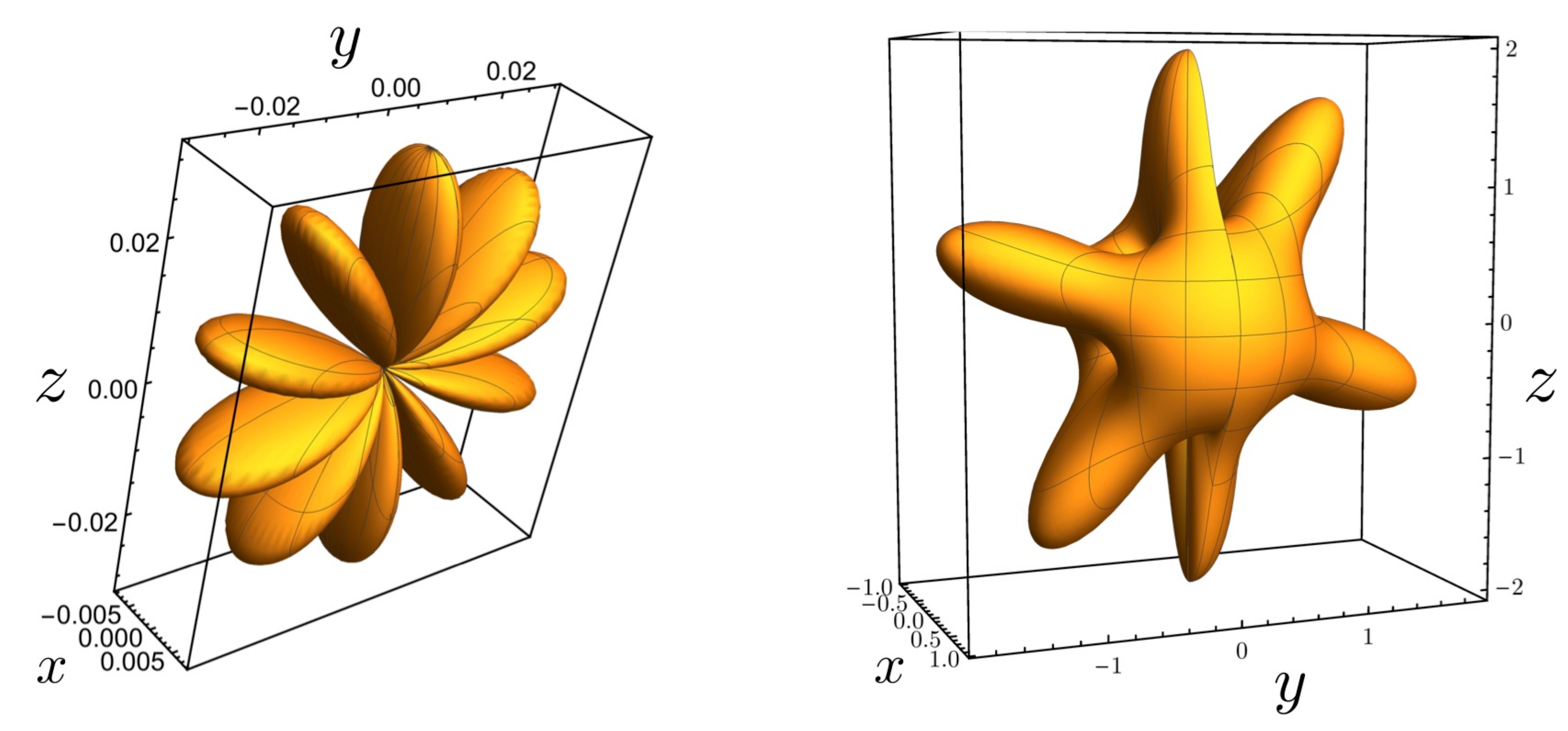}
\caption{Spherical polar plot of the interference term $-\text{Re}[q_+(5/2,\alpha,\beta)q^*_-(5/2,\alpha,\beta)]$ in the $Q_2(\alpha,\beta)$ function.}
\label{insieme}
\end{figure}

Using such expressions, we can compute $Q_j(\alpha,\beta)$ to investigate its features. However, looking at such function directly does not provide sufficient information for the discrimination of an incoherent mixture and a state such as $\ket{\psi-{1,2}}$. On the other hand, we find more informative to consider that $\frac12\left(|q_+(5/2,\alpha,\beta)|^2+|q_-(5/2,\alpha,\beta)|^2\right)$ is precisely the spherical SCS-based $Q$ function for the incoherent state $(\ket{S_1}\bra{S_1}\pm\ket{S_2}\bra{S_2})/2$. Let us call it $Q_{inc}(\alpha,\beta)$, so that 
\begin{equation}
Q_j(\alpha,\beta)=Q_{inc}(\alpha,\beta)+\text{sign}_j\text{Re}[q_+(5/2,\alpha,\beta)q^*_-(5/2,\alpha,\beta)],
\end{equation}
which pinpoints the contribution coming from the fixed-phase relation typical of a coherent superposition. We thus focus on state $\ket{\psi_2}$, which is the one that has been addressed in our experimental endeavors, and look at the term $-\text{Re}[q_+(5/2,\alpha,\beta)q^*_-(5/2,\alpha,\beta)]$, and represent it in the spherical polar plane defined above. Fig.~\ref{insieme} {\bf (a)} shows the results of our calculations. 

Such interference term exhibits 10 equally separated lobes, and is clearly displays both rotation and inversion symmetry. In fact, one can show that, for a generic value of $s$, the interference term in the corresponding $Q$ function exhibits $4s$ equally spaced lobes. It is worth mentioning that in Ref.~\cite{agarwalCAT1997} another figure of merit for the analysis of the effects of the interference term was adopted. More specifically, Ref.~\cite{agarwalCAT1997} studied the form of 
\begin{equation}
\frac{Q_j(\alpha,\beta)}{Q_{inc,j}(\alpha,\beta)}=1+\text{sign}_j\frac{2\text{Re}[q_+(5/2,\alpha,\beta)q^*_-(5/2,\alpha,\beta)]}{Q_{inc,j}(\alpha,\beta)},
\end{equation}
which thus quantifies the effect of quantum coherence as the deviation of $Q_j(\alpha,\beta)$ from $1$, whose representation in the chosen spherical polar space is a sphere of unit radius. When making use of such figure of merit, we find Fig.~\ref{insieme} {\bf (b)}, which shows a lobate behavior significantly different from the (incoherent) spherical trend. 

%
